# Gedanken Densities and Exact Constraints in Density Functional Theory

John P. Perdew[*,‡], Adrienn Ruzsinszky[*], Jianwei Sun[*], and Kieron Burke[**]

[*]Department of Physics, Temple University, Philadelphia, PA 19122

[‡]Department of Chemistry, Temple University, Philadelphia, PA 19122

[**]Department of Chemistry and Department of Physics, University of California, Irvine, CA 92697

**Abstract:** Approximations to the exact density functional for the exchange-correlation energy of a many-electron ground state can be constructed by satisfying constraints that are *universal*, i.e., valid for all electron densities. Gedanken densities are designed for the purpose of this construction, but need not be realistic. The uniform electron gas is an old gedanken density. Here, we propose a spherical two-electron gedanken density in which the dimensionless density gradient can be an arbitrary positive constant wherever the density is non-zero. The Lieb-Oxford lower bound on the exchange energy can be satisfied within a generalized gradient approximation (GGA) by bounding its enhancement factor or simplest GGA exchange-energy density. This enhancement-factor bound is well known to be sufficient, but our gedanken density shows that it is also necessary. The *conventional* exact exchange-energy density satisfies no such local bound, but energy densities are not unique, and the simplest GGA exchange-energy density is not an approximation to it. We further derive a strongly and optimally tightened bound on the exchange enhancement factor of a two-electron density, which is satisfied by the local density approximation but is violated by all published GGA's or meta-GGA's. Finally, some consequences of the non-uniform density-scaling behavior for the asymptotics of the exchange enhancement factor of a GGA or meta-GGA are given.

# 1. Gedanken densities: What and why? (Introduction and summary)

Kohn-Sham density functional theory [1,2] for the ground-state energy $E$ and density $n(\vec{r})$ of a many-electron system is widely used for atoms, molecules, and solids because of its balance between useful accuracy and computational efficiency. Only the exact density functional for the exchange-correlation energy needs to be approximated. The original local density (LDA) approximation [1,2] is

$$E_{xc}^{LDA}[n] = \int d^3r\, n(\vec{r}) \varepsilon_{xc}^{unif}(n(\vec{r})), \tag{1}$$

where $\varepsilon_{xc}^{unif}(n) = \varepsilon_{x}^{unif}(n) + \varepsilon_{c}^{unif}(n)$ is the exchange-correlation energy per electron of jellium, an electron gas of uniform density $n$. In the absence of a magnetic field, only the total density is needed in the exact theory, but the approximations work better under a generalization to spin-density functional theory [2], where they benefit from more input information. Here we will simplify the notation by displaying the approximations only for spin-unpolarized systems. For the exchange energy, which is our focus here, the generalization to arbitrary spin-polarization is trivial [3].

Jellium is not a real system, but it is one for which the LDA is exact. This is a gedanken density, in the same sense that an imagined experiment is a gedanken experiment. Jellium is also a quasi-realistic paradigm for the valence-electron density in a simple metal. The electron gas of non-uniform but slowly-varying density, on which the density-gradient expansion is based, is another gedanken density. Semilocal approximations can greatly increase the accuracy without much reducing the computational efficiency in comparison to LDA, and here we introduce new gedanken densities which may be useful for their further development. Model densities can also be interesting in other contexts, e.g., to explore exchange and correlation in systems of reduced dimensionality [4].

The simplest semilocal approximation is the generalized gradient approximation (GGA) [5-12]:

$$E_{xc}^{GGA}[n] = \int d^3r\, n\varepsilon_{x}^{unif}(n) F_{xc}^{GGA}(n,s), \tag{2}$$

where $\varepsilon_{x}^{unif}(n) = -(3/4\pi)(3\pi^2 n)^{1/3}$ is the exchange energy per electron of jellium and $F_{xc}^{GGA}(n,s) \geq 0$ is the enhancement factor over local exchange due to both correlation and the dimensionless density gradient

$$s = |\nabla n|/[2(3\pi^2)^{1/3} n^{4/3}] \ . \tag{3}$$

We can write

$$F_{xc}^{GGA}(n,s) = F_x^{GGA}(s) + F_c^{GGA}(n,s) \ . \tag{4}$$

The high-density limit of $F_{xc}^{GGA}(n,s)$ is the exchange enhancement factor $F_x^{GGA}(s)$, and $F_x^{GGA}(s=0) = F_x^{LDA} = 1$ is required to recover LDA in the uniform-density limit.

Unlike LDA, GGA is not unique. The nonempirical approach [2,6,9-12] to GGA construction requires the satisfaction of known exact constraints on the functional $E_{xc}[n]$ for *all* possible densities $n$. One of these constraints (automatically satisfied by LDA) is the Lieb-Oxford lower bound [13]. The original bound was for the indirect part of the electron-electron interaction, $<V_{ee}> - U[n]$ where $U[n]$ is the Hartree electrostatic interaction of the density with itself, for any wavefunction (not necessarily a ground-state) of density $n(\vec{r})$. Perdew [9] used the inequality $E_{xc}[n] \geq <V_{ee}> - U[n]$ to find a version that is more useful for density functional theory,

$$E_{xc}[n] \geq BE_x^{LDA}[n] = B\int d^3 r n \varepsilon_x^{unif}(n), \tag{5}$$

where of course $E_{xc}$ and $E_x$ are negative. The *optimal* (smallest possible) constant is in the range $1.67 < B < 2.273$. From the low-density limit for jellium, this range is narrowed [9] to $1.93 < B < 2.273$. Levy and Perdew [14] and recently Räsänen *et al.* [15] have conjectured that the optimum B is close to 1.93, but the only rigorous form of the bound has B=2.273 (or the slightly tighter 2.215 derived by Chan and Handy [16]). The right-hand side of Eq. (5) is clearly also a lower bound on $E_x[n]$ within density functional theory, since

$$E_x[n] \geq E_{xc}[n] \geq BE_x^{LDA}[n] \ . \tag{6}$$

Via the spin-scaling equality for exchange [3], *spin*-density GGA's are guaranteed to satisfy this rigorous bound on the exchange energy if the spin-unpolarized enhancement factor satisfies

$$F_x^{GGA}(s) < 2.273/2^{1/3} = 1.804 \quad \text{(all } s\text{, spin-unpolarized).} \tag{7}$$

In this form, the bound was used to construct the Perdew-Wang 1991 (PW91) [9-11] and the Perdew-Burke-Ernzerhof (PBE) [11,12] GGA's. In section 2, we will discuss the relevance of this and related newer exact constraints to functional construction.

While Eq. (7) is clearly a sufficient condition [12,17] for a GGA to satisfy the Lieb-Oxford bound of Eq. (6) on the exchange energy for all possible densities, is it also a necessary condition [17,18]? In other words, are GGA enhancement factors with $F_x^{GGA}(s) > 1.804$ over some range of $s$ strictly forbidden by the Lieb-Oxford lower bound? Refs. [17,19,20] (and the success of many functionals violating the bound) suggest that the answer is no.

Our answer to the last question is yes. To show this, in section 3 we construct a new gedanken two-electron spherical density in which the dimensionless density gradient $s$ takes the same arbitrary positive value over the entire density. The existence of such densities implies that GGA's with enhancement factors that go above 1.804 will violate the Lieb-Oxford bound for some of those densities.

By far the most important gedanken densities are the uniform or slowly-varying densities: an infinite set for which the computationally-efficient semilocal approximations can and should be exact, having some resemblance to real densities of interest (those of valence electrons in simple metals). Our new gedanken density of section 3 is less realistic, and serves a more limited purpose: to establish that bounds like Eqs. (7) or (11) are necessary constraints on semilocal approximations. We would like to find another gedanken density, in which $\alpha$ is everywhere infinite, but have not found one yet.

An early [20] and recent [19] objection to the inequality of Eq. (7) was that it is contradicted by the behavior of the conventional exact exchange energy density in the density tail of an atom or molecule. But this objection overlooks the fact that the exact exchange energy density is not unique, and that the simplest GGA exchange-energy density does not (and cannot) model the conventional choice. We discuss this issue further in section 4, where we also present non-conventional exact exchange energy densities with exact exchange enhancement factors that are (at least for a cuspless two-electron atom) everywhere bounded from above by 1.804.

It is now well-known that the GGA form is too simple to be highly accurate, even for those densities for which semilocal approximations are well-suited. Much better dissociation energies, surface energies, and equilibrium geometries [21] can be found from the semilocal (hence still computationally efficient) meta-GGA [21-26], which uses three ingredients: the local density $n(\vec{r})$, its gradient $\nabla n(\vec{r})$, and the orbital kinetic energy density

$$\tau(\vec{r}) = 2\sum_{\alpha}^{occup} |\nabla \psi_\alpha(\vec{r})|^2 / 2 \quad \text{(spin-unpolarized)}, \tag{8}$$

where $\psi_\alpha(\vec{r})$ is a Kohn-Sham orbital. Ref. [26] suggests that a meta-GGA

$$E_{xc}^{MGGA}[n] = \int d^3 r n \varepsilon_x^{unif}(n) F_{xc}^{MGGA}(n, s, \alpha), \tag{9}$$

with

$$\alpha = (\tau - \tau^W) / \tau^{unif}, \tag{10}$$

can recognize and assign a different appropriate GGA description to covalent single ($\alpha \approx 0$), metallic ($\alpha \approx 1$), and weak ($\alpha \gg 1$) bonds. Here the Weizsäcker expression $\tau^W = |\nabla n|^2 / 8n$ is exact for a two-electron density ($\alpha = 0$) and $\tau^{unif} = (3/10)(3\pi^2)^{2/3} n^{5/3}$ is exact for a uniform density ($s = 0, \alpha = 1$). $\alpha$ is the sole ingredient of the Becke-Edgecombe [27] electron localization function (ELF), but here we think of it as a dimensionless deviation from single orbital shape (DDSOS).

The high-density limit of the meta-GGA enhancement factor $F_{xc}^{MGGA}(n, s, \alpha) \geq 0$ is again the exchange enhancement factor $F_x^{MGGA}(s, \alpha)$ with $F_x^{MGGA}(s = 0, \alpha = 1) = F_x^{LDA} = 1$, and again the sufficient and necessary condition to satisfy the global Lieb-Oxford bound on the exchange energy for all possible densities with $\alpha = 0$ is $F_x^{MGGA}(s, \alpha = 0) < 1.804$. But, in section 5, we will derive the much tighter bound

$$F_x^{MGGA}(s, \alpha = 0) < 1.174 \quad \text{(all } s\text{, spin-unpolarized)} \tag{11}$$

and conjecture that this tight bound for meta-GGA exchange might remain true for all $\alpha$.

Eq. (11) is strongly violated by existing GGA's and meta-GGA's. For the PBE GGA [12], $F_x^{GGA}(s)$ varies from $1 + 0.2195 s^2 + ...$ at small $s$ to 1.804 at large $s$. For the revTPSS meta-GGA [24], $F_x^{MGGA}(s, \alpha = 1) = 1 + 0.1235 s^2 + \cdots$ and $F_x^{MGGA}(s, \alpha = 0) = 1.15 + 0 s^2 + \cdots$ at small $s$ while $F_x^{MGGA}(s, \alpha) = 1.804$ at large $s$. These functionals were constructed to satisfy Eq. (7), not Eq. (11).

In section 6, we discuss non-uniform density-scaling of the exchange energy to the true two-dimensional limit. We argue that, to satisfy the right scaling behavior, we must have

$$\lim_{s \to \infty} F_x^{GGA}(s) \propto s^{-1/2}, \tag{12}$$

and

$$\lim_{s \to \infty} F_x^{MGGA}(s, \alpha = 0) \propto s^{-1/2}. \tag{13}$$

## 2. Relevance of these constraints to functional construction

It can be argued that many approximate exchange functionals (e.g., B88 [7]) violate Eq. (7), that nearly all violate Eq. (13), and that so far all beyond LDA violate Eq. (11). Many of these functionals are usefully accurate for real systems. So, are these constraints merely pedantic? We will argue that the answer is no, from several different perspectives.

It is precisely because we want to refine the functionals, and especially the meta-GGA's, that we now focus on exact constraints to which many properties of real systems are not so sensitive. The satisfaction of exact constraints can subtly improve a functional for a given system by controlling the way the functional approaches extreme limits, even when the system is not close to those extremes. Exact constraints that are satisfied by LDA, such as Eq. (11), are of special theoretical interest [28,29,30], and should be preserved in beyond-LDA functionals.

The exact density functional and its exact constraints are universal, for all allowable densities, and not just for the densities of real atoms, molecules, and solids. While the latter systems have great practical interest, their more familiar properties may not sensitively sample all the exact constraints. If the form of a functional (e.g., GGA or meta-GGA) permits the satisfaction of an exact constraint for all densities, then that constraint should be taken seriously in the nonempirical construction of functionals of that form.

Even for the known properties of real systems, the existing functionals of any beyond-LDA form are far from optimal. The GGA form can satisfy a certain set of constraints, but cannot satisfy them all simultaneously [31,32]. GGA's that work best for the atomization energies of molecules (like B88) tend to overestimate the lattice constants of solids, while those that predict accurate lattice constants (like PBEsol [32]) strongly overestimate atomization energies [32]. Meta-GGA's often resolve these dilemmas [21,23,24], but can still fail when confronted with properties for which they were not previously tested, such as the critical pressures for structural phase transitions in solids [33]. Similar considerations may apply to small energy differences between different structures of a molecule. We doubt that an empirical approach to functional construction can deal with such problems.

The semilocal functionals such as meta-GGA are most appropriate to certain classes of densities: (1) uniform or slowly-varying densities, for which they can be exact by construction, and (2) compact densities, such as the densities of atoms, where the exact exchange and correlation holes are necessarily confined to a region close to the electron they surround. For such densities, the meta-GGA exchange and meta-GGA correlation energies can be separately accurate. For multi-center bonded systems, such as molecules and solids, the exact exchange hole and the exact correlation hole can be separately spread out over two or more centers. In many of those cases (e.g., near the equilibrium geometries of *sp*-bonded systems), the exact combined exchange-correlation hole is still localized around the electron it surrounds, and semilocal functionals can work via a cancellation of errors between exchange and correlation. (This error cancellation would be lost if we combined exact exchange with semilocal correlation.) Although an error cancellation is expected on the basis of strong qualitative arguments, the quantitative degree of cancellation is not something that can be predicted or controlled. Thus even a meta-GGA that satisfies all possible exact constraints is still not guaranteed to work; testing and benchmarking are always needed.

Going beyond the meta-GGA form is necessary for certain strongly-correlated or stretched-bond situations. Some of the fully nonlocal approximations, such as local hybrid functionals [34] or self-interaction corrections [35,36], require a good meta-GGA as a starting point. The self-interaction correction in particular requires applying the meta-GGA to the densities of typically localized orbitals, constructed by unitary transformation of the occupied Kohn-Sham orbitals. These one-electron ($\alpha = 0$) densities are somewhat more challenging to a meta-GGA than the ground-state density of the spin-polarized one-electron atom. We expect that satisfaction of Eqs. (11) and (13), along with more familiar constraints, could greatly improve the accuracy of self-interaction-corrected results.

Finally, we note that a functional can easily satisfy a bound like Eq. (6) for all realistic densities, and still violate it for some allowable densities. The gedanken density of the next section suggests that, to satisfy such a bound for *all* allowable densities, requires a corresponding bound like $F_x < B$ on the enhancement factor.

## 3. A gedanken two-electron spherical density with constant non-zero *s*

Here we will introduce a new gedanken density to show that the bound of Eqs. (7) or (11) is a *necessary* condition for a GGA or meta-GGA to satisfy the corresponding lower bound on the integrated exchange energy for *all* possible electron densities.

Consider the density
$$n(r) = A/r^3 \quad (R_0 < r < R_1) \tag{14}$$
$$0 \text{ (otherwise)}.$$

Here $A$, $R_0$, and $R_1 > R_0$ are positive constants, and the density is normalized to two electrons:

$$\int_{R_0}^{R_1} dr 4\pi r^2 A/r^3 = 4\pi A \ln y = 2, \tag{15}$$

which fixes $A$ as a function of $y = R_1/R_0$. (Other electron numbers are also possible, but for *N*=2 it is easy to evaluate the exact exchange energy.) The reduced gradient is then

$$s = (3/2)[(2/\{3\pi\}) \ln y]^{1/3}, \tag{16}$$

the same positive but arbitrary constant over the whole range where the density is non-zero. When $y$ varies from 1 to $\infty$, $s$ varies from 0 to $\infty$ (Fig. 1). Clearly, when *s* is constant where the density is non-zero, the enhancement factor $F_x^{GGA}(s)$ can be factored out of the integral for $E_x^{GGA}$. Then that GGA will satisfy the global exchange-only Lieb-Oxford bound for this family of densities only when $F_x^{GGA}(s) < 1.804$ for *all s*.

For this gedanken density, we can evaluate $E_x^{LDA}$ and the exact $E_x$ (which for two-electron ground-state densities equals $(-1/2)U[n]$), finding

$$E_x^{LDA} = [-3(3\pi^2)^{1/3}/\{(2\pi)^{4/3} R_0\}][1 - 1/y]/[\ln y]^{4/3}, \tag{17}$$

$$E_x = (-2/R_0)[1-(1+\ln y)/y]/[\ln y]^2, \tag{18}$$

where $y = R_1/R_0$. The ratio $E_x/E_x^{LDA}$ (Fig. 2) maximizes at 1.0875 for $y \approx 5.3$ ($s \approx 1.06$), and tends slowly to zero in the $y \to 1$ ($s \to 0$) and $y \to \infty$ ($s \to \infty$) limits. This result is consistent with Eq. (11). The gedanken density of Eq. (14) in the limit $y \to 1$ becomes a thin spherical shell of high uniform two-electron density, for which no general-purpose semilocal approximation could be expected to work.

A possible objection is that this density is too sharply cut off to be allowable, i.e, to arise from a potential. However, we can make allowable densities by introducing regions $R_0 - \Delta/2 < r < R_0 + \Delta/2$ and $R_1 - \Delta/2 < r < R_1 + \Delta/2$ over which the density is rounded to approach zero with zero slope at $R_0 - \Delta/2$ and $R_1 + \Delta/2$. $\Delta$ can be any positive length less than $2R_0$. If a GGA violates Eq. (7), there will always be a range of $s$ or $y$ for which this GGA violates the global Lieb-Oxford bound for allowable rounded densities with small-enough $\Delta$. This is true regardless of what happens in the rounding regions, from which the contributions to the GGA exchange energy must be negative.

As $\Delta$ approaches zero from above, the reduced gradient $s$ becomes very large over the small region where the rounding occurs. But, for a GGA which satisfies Eq. (7), the contribution to $E_x^{GGA}$ from the rounding regions must vanish as fast as or faster than $\Delta$. For an even smoother rounding leading to the same conclusions, see Appendix A.

The orbital kinetic energy density $\tau = \tau^W$ diverges like $s^2$ as $s \to \infty$, and $s \propto 1/\Delta$ in the rounding region as $\Delta \to 0$. Thus the kinetic energy diverges like $(1/\Delta)^2 \Delta = 1/\Delta$ when $\Delta \to 0$. Eq. (14) is not a Lieb allowable [37] density.

In summary, our gedanken density of Eq. (14) is not an allowable density, but it is the limit of a sequence of allowable densities associated with a convergent sequence of approximate exchange energies and bounds. Our approximations should obey the Lieb-Oxford bound on the exchange energy for every allowable density in the sequence.

The Becke 1988 [7] GGA has an enhancement factor $F_x^{B88}(s)$ that grows without bound as $s$ increases. It will violate the global Lieb-Oxford bound on $E_x[n]$ for all $y = R_1/R_0 > \exp(4\pi s_0^3/9) = 2x10^{19}$ where $F_x^{B88}(s_0 = 3.17) = 1.804$. While our gedanken density of Eq. (14) does not so far present a serious practical challenge even to Becke 1988 exchange, it will present a much more serious challenge after the bound of Eq. (7) is tightened to that of Eq. (11) in section 5.

Here we briefly mention an earlier gedanken density [14] that gave us some limited guidance for the 1996 construction of the PBE GGA: Consider a one-electron density that is lattice-periodic over a large crystal. As the volume $\Omega$ of the crystal tends to infinity, the reduced gradient $s$ tends to infinity everywhere like $\Omega^{1/3}$. A GGA whose enhancement factor $F_x^{GGA}(s)$ exceeds 1.804 in the limit $s \to \infty$ will violate the general Lieb-Oxford bound for this density.

## 4. Energy densities and GGA enhancement factors

In Eq. (2), $n\varepsilon_x^{unif}(n)F_{xc}^{GGA}(n,s)$ is a function of position $\vec{r}$ that can be interpreted as a GGA exchange-correlation energy density, and $n\varepsilon_x^{unif}(n)F_x^{GGA}(s)$ plays the same role for exchange. The best-known exact exchange energy density is the conventional one [38]

$$n(\vec{r})\varepsilon_x^{conv}(\vec{r}) = (-1/4)\int d^3r' |\rho(\vec{r},\vec{r}')|^2 / |\vec{r}'-\vec{r}| \quad \text{(spin-unpolarized)}, \tag{19}$$

where $\rho(\vec{r},\vec{r}') = 2\sum_\alpha^{occup} \psi_\alpha^*(\vec{r})\psi_\alpha(\vec{r}')$ is the Kohn-Sham one-particle density matrix. Then it may be tempting to equate

$$F_x^{GGA} = \varepsilon_x^{conv} / \varepsilon_x^{unif}, \tag{20}$$

as in Ref. [19] or in the earlier Ref. [20]. In these references, the right-hand side of Eq. (20) was evaluated numerically for atoms and molecules, and found to diverge in the tail of the electron density. That unbounded result, which contradicts Eq. (7), can also be found analytically, since in the tail $\varepsilon_x^{conv} \to -1/(2r)$ [7] while $\varepsilon_x^{unif}$ decays exponentially with $r$. The right-hand side of Eq. (20), plotted vs. $s$ in molecules, presents a band of values [19] rather than a single value for each $s$.

Energy densities are not observables, and are not uniquely defined in density functional theory (with the electron gas of uniform density as the sole exception). One can add any function that integrates to zero to any choice of energy density, and the integral will not change; thus this produces another equally valid energy density. This fact is very well known in studies of the non-interacting kinetic energy $T_s$. There are two natural choices of kinetic energy density, which both integrate to $T_s$: The first is the positive $\tau$ of Eq. (8), and the second is

$$\tilde{\tau} = 2\sum_\alpha^{occup} \psi_\alpha^*(\vec{r})(-1/2)\nabla^2 \psi_\alpha(\vec{r}). \tag{21}$$

Each can be useful in its own context, but they differ substantially from each other. Weighted sums like $c\tau + (1-c)\tilde{\tau}$ are also possible choices. It is precisely for this reason that the choice of kinetic energy density must be specified when meta-GGA's are being constructed.

The same reasoning applies to exchange or exchange-correlation energy densities. They are not uniquely defined. For example, $n\varepsilon_x^{conv}$ and $n\varepsilon_x^{conv} + c\nabla^2 n^{2/3}$ integrate to the same exchange energy. There is no reason to equate the simplest GGA exchange energy density to $n\varepsilon_x^{conv}$, as in Eq. (20). Indeed, they should not be equated, because the second-order gradient expansion of the conventional exchange energy density involves ill-behaved terms that cannot be expressed in terms of $n$ and $s$ alone [39,40]. An integration by parts, which changes the conventional exchange energy density to an unconventional one, must be performed to express the gradient expansion in terms of the latter two variables alone [41]. Several years ago, exchange energy densities were defined [42] in terms of the exchange potential. Such energy densities are unambiguously defined for any exchange energy functional, exact or approximate, but their interpretation and use is too demanding at present.

This situation is familiar in other areas of physics. The scalar and vector potentials of electromagnetic theory [43] are not measurable, and neither is the wavefunction of quantum mechanics [44], so these objects are gauge or unitarily variant, while measurable properties are gauge or unitarily invariant.

We have never asserted the existence of exact exchange energy densities satisfying a bound like that of Eq. (7), but we close this section by presenting suggestive but inconclusive evidence that such exact exchange energy densities might exist:

A family of exact exchange energy densities can be generated by coordinate transformation [38,45-47]:

$$n(\vec{r})\varepsilon_x^\lambda(\vec{r}) = (-1/4)\int d^3u \left|\rho(\vec{r}+[\lambda-1]\vec{u}, \vec{r}+\lambda\vec{u})\right|^2 / u. \tag{22}$$

When the parameter $\lambda = 1$, the conventional Eq. (19) is recovered, with $\varepsilon_x^{\lambda=1} \to -1/(2r)$ in the density tail for an atom or molecule. Any $\lambda$ between 0 and 1 makes $\varepsilon_x^\lambda$ decay exponentially for an atom or molecule, with the fastest decay for $\lambda = 1/2$ [38,45]. Note that any member of this family can be as easily constructed from a knowledge of the orbitals as can the conventional choice ($\lambda = 1$), and is an equally valid choice for an exact exchange energy density. Now let us define an effective, $\lambda$-dependent exchange enhancement factor

$$F_x^\lambda = \varepsilon_x^\lambda / \varepsilon_x^{unif} . \tag{23}$$

We have evaluated Eq. (23) and plotted it vs. $\beta r$ for a simple cuspless two-electron atom of density

$$n(r) = (1/2\pi)\beta^3(1+2\beta r)\exp(-2\beta r), \tag{24}$$

for which $s$ is a function of $\beta r$, increasing from zero at $\beta r = 0$ to $\infty$ as $\beta r \to \infty$. (Our numerical tests confirm that the integrated exact exchange energy $-0.492\beta$ is independent of $\lambda$.) Our results (independent of the scale parameter $\beta$) are plotted in Fig. 3. While $F_x^\lambda$ diverges in the density tail for $\lambda = 1$, it everywhere satisfies the bound of Eq. (7) for $0.75 > \lambda > 0.50$. At $\lambda = 0.75$, it tends to a positive constant as $\beta r \to \infty$.

Finally, note that it is the *integrated* exchange and correlation energies that are to be approximated to satisfy exact constraints. The associated energy densities are not relevant to experiment, and not relevant to theory except for example in the construction of local hybrid functionals [34]. While there are many possible exchange energy densities for a given GGA, some more bounded than others, there is only one of the simple form $F_x^{GGA}(s)n\varepsilon_x^{unif}(n)$, and that one should satisfy the bound of Eq. (6).

## 5. Tight bound on the exchange-enhancement factor for a two-electron density

The Lieb-Oxford bound of Eq. (6) is valid for any density, but Lieb and Oxford [13] also derived tighter bounds for one- and two-electron densities. Their derivation of Eq. (25) below was presented as a more rigorous version of an earlier one by Gadre, Bartolotti, and Handy [48]. Because the GGA form cannot distinguish between a two-electron density and one with more electrons, enforcement of a tight Lieb-Oxford bound within GGA would lead to little improvement over LDA for most systems. Thus this is an exact constraint that cannot be usefully imposed on GGA construction, but can be very useful for meta-GGA's.

For an arbitrary spin-polarized one-electron density $n_1(\vec{r})$, where $<V_{ee}> - U[n] = E_{xc} = E_x$ is a pure self-interaction correction, the *optimal* bound is known [13,48]:

$$E_x[n_1] \geq -1.092 \int d^3 r \, n_1^{4/3} . \tag{25}$$

For a spin-unpolarized two-electron ground state of density $n_2$, we can take $n_1 = n_2/2$ and $E_x[n_2] = 2E_x[n_1]$. Then Eq. (25) implies

$$E_x[n_2] \geq (1.092/2^{1/3})[4\pi/\{3(3\pi^2)^{1/3}\}] E_x^{LDA}[n_2] = 1.174 E_x^{LDA}[n_2] . \tag{26}$$

Two-electron ground states have $\alpha = 0$. Thus our two-electron gedanken density of section 3 tells us that a sufficient and necessary condition for a meta-GGA to satisfy Eq. (26) for all two-electron densities is Eq. (11),

$$F_x^{MGGA}(s, \alpha = 0) \leq 1.174 \quad \text{(all } s\text{, spin-unpolarized)}. \tag{27}$$

Eq. (27) remains optimally tight, as it would not be if it were derived from the Lieb-Oxford bound on $<V_{ee}> - U[n]$ for a two-electron density. We have no proof of the analog of Eq. (27) for $\alpha > 0$, but we suspect that it may be true, because we suspect that $F_x^{MGGA}(s, \alpha) < F_x^{MGGA}(s, \alpha = 0)$ as in the Meta-GGA Made Simple of Ref. [25] (which works rather well with PBE GGA correlation and satisfies $F_x^{MGGA}(s, \alpha) \leq 1.29$). We know of no two-electron spin-unpolarized density with $E_x / E_x^{LDA} > 1.174$.

Capelle and Odashima [49] have suggested the possibility of a tightened Lieb-Oxford bound for the exchange-correlation energy, and we suspect that they were right to do so. However, the possibilities for tightening this bound are limited, reducing *B* of Eq. (5) from 2.273 to a value 1.93 or higher (and thus the bound on the spin-unpolarized $F_x$ from 1.804 to 1.53), as mentioned after Eq. (5). In contrast, our suspected bound $F_x^{MGGA}(s, \alpha) < 1.174$ is dramatically tightened over Eq. (7). We will explore the consequences of this assumption in future work.

While LDA satisfies Eq. (27), published GGA's and meta-GGA's violate it. So why do GGA's work even for the exchange energy of the He atom and similar two-electron densities? The answer must be that, for these two-electron densities, GGA's make an error cancellation between regions of small s (where their exchange enhancement is too low, around 1, as it must be to recover the uniform-density limit) and larger s (where their exchange enhancement is too high, violating Eq. (27)). The Meta-GGA Made Simple [25], and to a lesser extent other meta-GGA's, have $F_x^{MGGA}(s \approx 0, \alpha = 0)$ considerably higher than 1 but less than 1.174 (see Fig. 1 of Ref. [25]). The bound of Eq. (27) cannot be applied at the GGA level, even for the He atom, because it would destroy this error cancellation, but it can hopefully be applied at the meta-GGA level. Many-electron densities on the other hand are energetically dominated by regions with $s \leq 1$, where standard functionals are not seriously challenged even by our tightened lower bound.

A meta-GGA for exchange that satisfies the conjectured general bound $E_x^{MGGA}[n] > 1.174 E_x^{LDA}[n]$ might work with semilocal (sl) functionals for correlation, which typically satisfy $E_c^{sl}[n] > 0.94 E_x^{LDA}[n]$, making $E_{xc}^{sl}[n] > 2.14 E_x^{LDA}[n]$.

## 6. Non-uniform density scaling: Implications for the asymptotics of the exchange enhancement factor

In this section, we will explore the implications of non-uniform density scaling [50] for the large-$s$ and large-$\alpha$ behaviors of the exchange enhancement factor $F_x$. We start from a density $n(x, y, z)$ having a finite $E_x[n]$, then define the one-dimensionally scaled density $n_\lambda^{(1)}(x, y, z) = \lambda n(\lambda x, y, z)$. The scaled density has the same electron number as the unscaled one, but is more compressed ($\lambda > 1$) or expanded ($\lambda < 1$) along the x direction. When $\lambda \to \infty$, the density collapses from three dimensions to two, in which the exchange energy is still finite and negative. Levy [50] proved that

$$\lim_{\lambda \to \infty} E_x[n_\lambda^{(1)}] > -\infty. \tag{28}$$

The LDA and most existing GGA's and meta-GGA's violate Eq. (28) [51-53]. The PW91 GGA [9-11] and the VT(8,4) GGA [54] and its related meta-GGA [55] satisfy Eq. (28), but they incorrectly [52,53] make the left-hand side vanish. A finite limit is achieved by the GGA of Chiodo et al. [56].

Starting from the definitions of s (Eq. (3)) and $\alpha$ (Eq. (10), we easily find that, under non-uniform scaling to the true two-dimensional limit ($\lambda \to \infty$),

$$s(x, y, z) \to \lambda^{2/3} f(\lambda x, y, z), \tag{29}$$

$$\alpha(x, y, z) \to 0. \tag{30}$$

If the unscaled $s$ is nonzero over some region in which the unscaled density is non-zero, then the meta-GGA exchange energy (Eq. (9)) has a finite non-zero limit for Eq. (28) when Eq. (13) is satisfied. This determines how the exchange enhancement factor vanishes as $s \to \infty$.

Levy [50] also defined a two-dimensional scaling of $n(x, y, z)$ to

$$n_\lambda^{(2)}(x, y, z) = \lambda^2 n(x, \lambda y, \lambda z). \tag{31}$$

Clearly this is the product of two one-dimensional scalings along different axes with the same scale parameter. Applying a third yields the three-dimensional or uniform scaling

$$n_\lambda^{(3)}(x, y, z) = \lambda^3 n(\lambda x, \lambda y, \lambda z). \tag{32}$$

Our exchange functionals and essentially all other sensible exchange functionals are designed to behave correctly [57] under uniform scaling:

$$\lambda^{-1} E_x[n_\lambda^{(3)}] = E_x[n]. \tag{33}$$

Moreover, the exchange component of a sensible approximate exchange-correlation energy functional is found when the density is scaled uniformly by $\lambda$, the functional is divided by $\lambda$, and $\lambda$ is taken to $\infty$, because this is a known property of the exact exchange-correlation functional [58].

Functionals that satisfy both Eq. (28) and Eq. (33) will of course also scale correctly [50] under two-dimensional scaling to the one-dimensional limit:

$$\lim_{\lambda \to \infty} \lambda^{-1} E_x[n_\lambda^{(2)}] > -\infty. \tag{34}$$

In the low-density limit under non-uniform three-dimensional scaling, correlation scales like exchange. We can define

$$B_{xc}[n] = \lim_{\lambda \to 0} \lambda^{-1} E_{xc}[n_\lambda^{(3)}], \tag{35}$$

which itself has one-dimensional and two-dimensional scaling limits that will be satisfied by a GGA (or in generalization by a meta-GGA) if [14]

$$\lim_{s \to \infty} s^{1/2} F_{xc}^{GGA}(n, s) < \infty. \tag{36}$$

When the exchange part of $F_{xc}^{GGA}$ satisfies Eq. (13), it is very easy to satisfy Eq. (36).

**Notes added in proof:** Mirtschink, Seidl, and Gori-Giorgi [59] have discussed the "local Lieb-Oxford bound". Peverati and Truhlar [60] have written that further improvement of approximate density functionals "may involve breaking some of the constraints that we are following right now or – less likely as it seems to us – adding new constraints". We hope that the new constraints of our paper will lead to practical improvements.

**Acknowledgments:** Our work on exact constraints has been inspired by that of Mel Levy. The work of JPP and JS was supported in part by the National Science Foundation under grant DMR-1305135. That of KB was supported by DOE under grant DE-FG02-08ER46496. We thank G.I. Csonka for assistance with the manuscript, and S.B. Trickey as well as our referees for comments and corrections.

**Appendix: Well-behaved densities tending to the gedanken density of Eq. (14)**

In this appendix, we check that the density of Eq. (14) can be achieved as a limit of densities that satisfy physically reasonable physical conditions:

$$\infty > n(\mathbf{r}) > 0 \tag{A1}$$

everywhere in real-space, $n(\mathbf{r})$ is normalizable, and $n(\mathbf{r})$ has finite non-interacting kinetic energy [37]. We will also require that its second derivative be finite everywhere, so as to be able to find the corresponding Kohn-Sham potential (i.e., the density is non-interacting $v$-representable.)

To construct such a density from that of Eq. (14), we begin by extending the density to be finite in all regions of space. We write

$$n_e(\mathbf{r}) = \frac{A}{R_0^3}\left[10 - 15\frac{r}{R_0} + 6\left(\frac{r}{R_0}\right)^2\right], r \leq R_0$$
$$= A/r^3, r \geq R_0, \tag{A2}$$

where the form for $r < R$ is a quadratic chosen to match the gedanken density and its first two derivatives at $R_0$. Note that $n_e(\mathbf{r})$ coincides with the gedanken density exactly between $R_0$ and $R_1$, but remains finite everywhere.

Next, we multiply by a damping factor to ensure that the density drops rapidly outside the shell. We define the damping function

$$f_m(x) = \frac{\left(\sum_{j=0}^{m} \frac{x^j}{j!}\right)\exp(-x)}{\sqrt{1 + \left(\frac{x^m}{m!}\right)^2}}, \quad x \geq 0$$
$$= 1, x \leq 0 \quad , \tag{A3}$$

which switches from a constant (1) to a decaying exponential at $x = 0$. Here $m$ determines the number of derivatives that remain continuous at $x = 0$, while the denominator ensures that a simple exponential decay is recovered at large $x$. We choose $m = 2$ to ensure that second-derivatives are well-behaved at the transition.

Our well-behaved density can now be defined as

$$n_\gamma(\mathbf{r}) = n_e(\mathbf{r}) f_2\left(\frac{R_0 - r}{\gamma} + \frac{1}{2}\right) f_2\left(\frac{r - R_1}{\gamma} + \frac{1}{2}\right) \tag{A4}$$

and as $\gamma \to 0$, it approaches the gedanken density of the text. To ensure normalization, $A$ must become a function of $\gamma$ whose $\gamma \to 0$ limit is that of Eq. (15) of the text. For any finite $\gamma$, this is a Lieb-allowed density, and for sufficiently small $\gamma$, its exchange energy can be made arbitrarily close to the gedanken density of the main text.

In Fig 4, we plot densities for $R_0 = 1$, and $R_1 = 2$ with $\gamma = 0.1$, as well as the gedanken density. Clearly our density is well-behaved, and matches (up to the normalization constant) the gedanken density for $R_0 + \frac{\gamma}{2} < r < R_1 - \frac{\gamma}{2}$.

In Fig 5, we plot $s(r)$ for our density, noting its constant value in the interior of the shell, although it becomes very large outside in the exponentially decaying regions. In fact, $s$ diverges as $r \to \infty$, just as in real atoms, but the density decays exponentially.

In Fig 6, we plot $v_s(r)$ for this two-electron density, with the constant chosen to make $v_s$ vanish as $r \to \infty$. The eigenvalue $\varepsilon$ is $-1/(8\gamma^2)$, which is -12.5 for $\gamma$

=0.1. The KS potential is continuous everywhere by construction. For the gedanken density, $v_s(r) = \varepsilon + 3/(8r^2)$ inside the shell.

Fig. 1: Plot of the reduced density gradient $s$ (Eq. (16)) vs. $y = R_1/R_0$ for the two-electron gedanken density of Eq. (14).

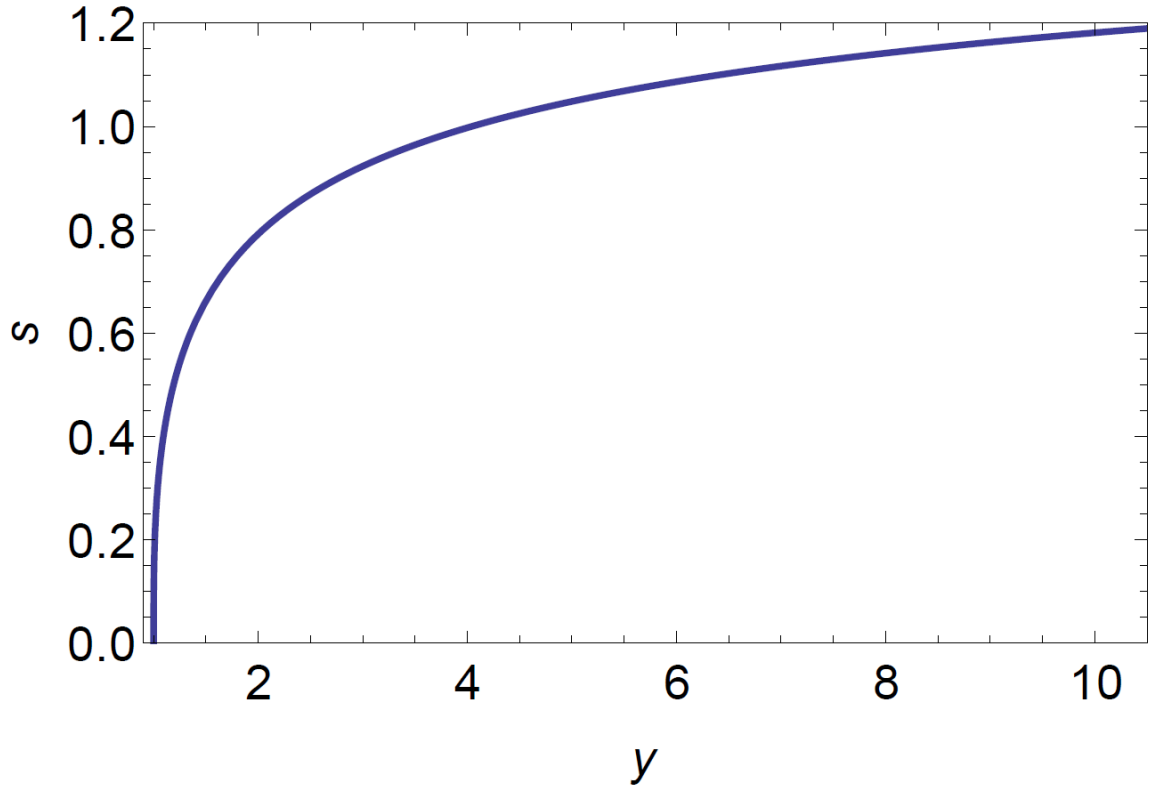

Fig. 2: Plot of the exchange-energy ratio $E_x / E_x^{LDA}$ (Eqs. (17) and (18)) vs. $y = R_1 / R_0$ for the two-electron gedanken density of Eq. (14).

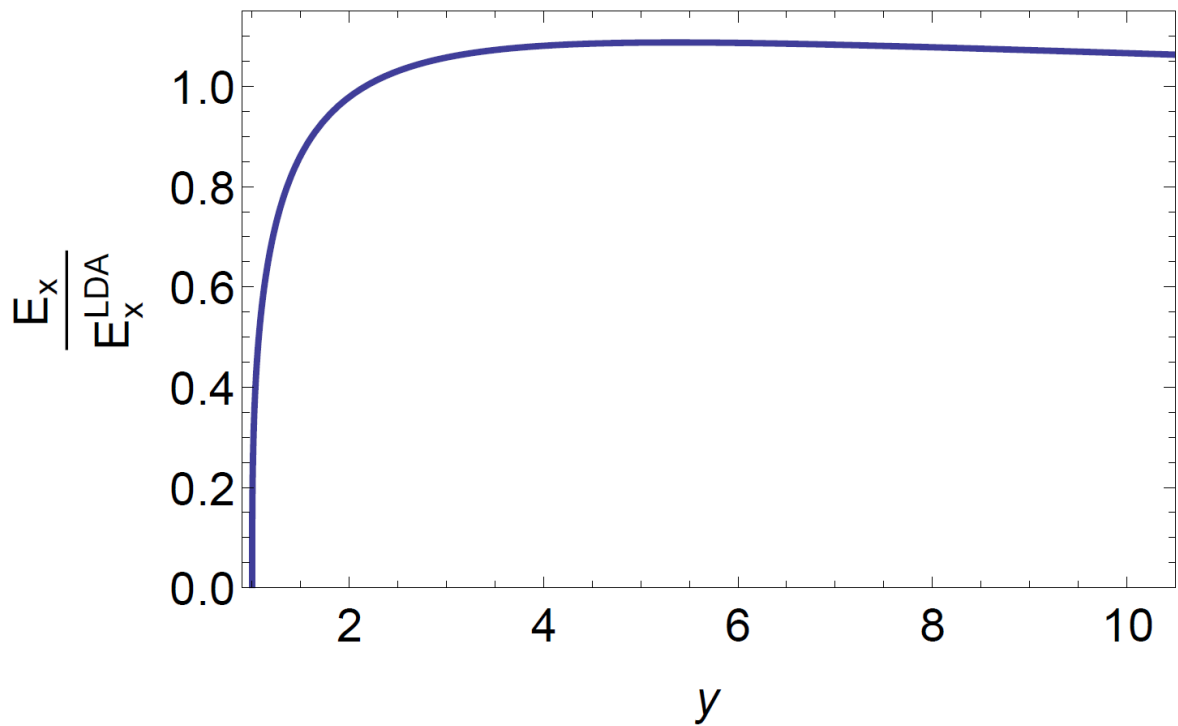

Fig. 3: Plot of the effective $\lambda$-dependent exchange enhancement factor of Eq. (23) vs. $\beta r$ for the cuspless two-electron ($\alpha = 0$) density of Eq. (24). Different values of $\lambda$ correspond to different choices for the exact exchange energy density under a coordinate transformation. The curve for $\lambda = 0.75$ tends to a nonzero constant as $\beta r \to \infty$. All curves in $0.75 > \lambda > 0.5$ satisfy the bound of Eq. (7). Note that $s =$ 0.00, 0.54, 1.06, 1.98, and 3.67 at $\beta r = 0, 1, 2, 3,$ and 4, respectively.

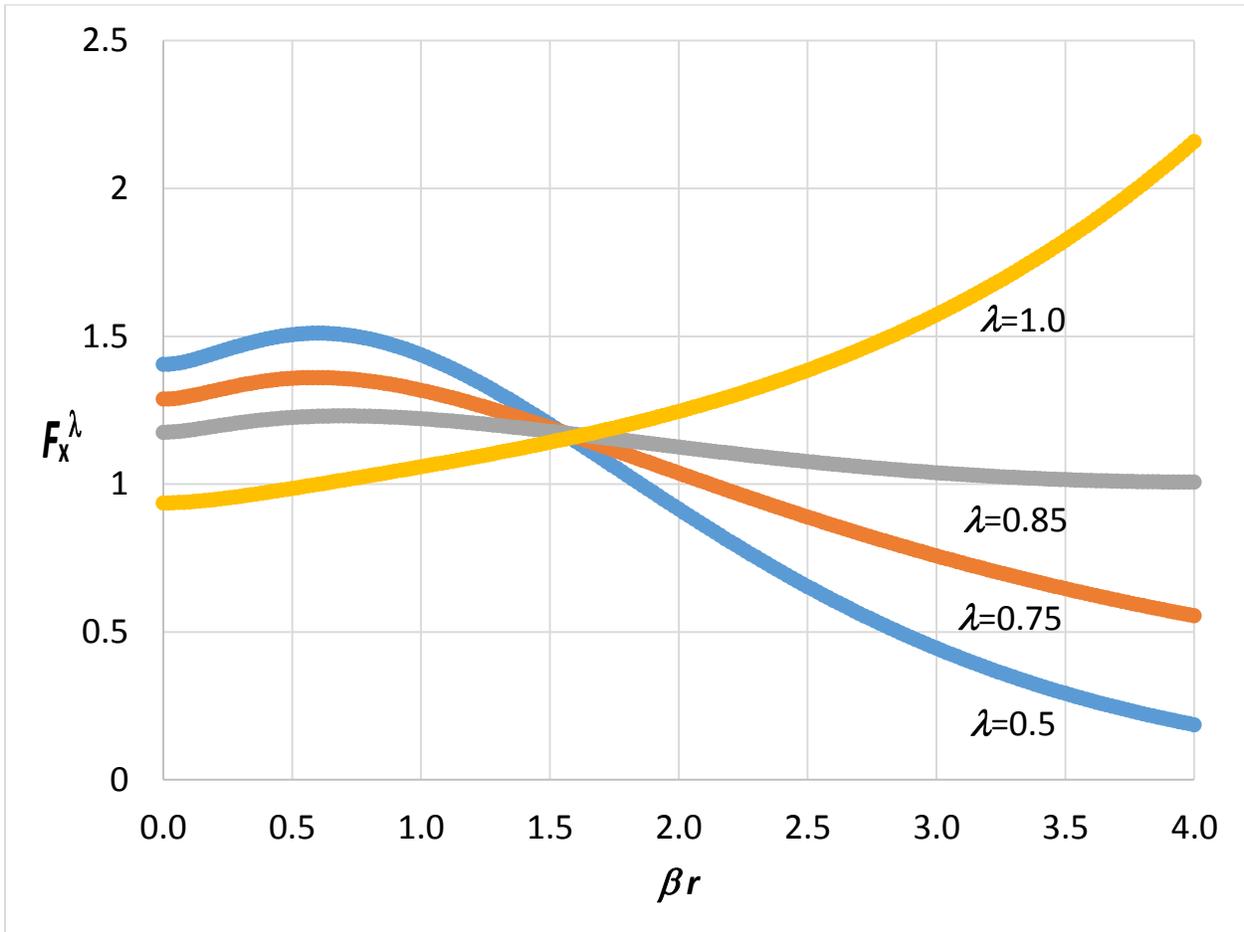

Fig. 4: Plot of the smoothed two-electron gedanken density $n$ of Eq. (A4) vs. distance $r$ from the origin. The $\gamma = 0.001$ curve already converges to the gedanken density of Eq. (14).

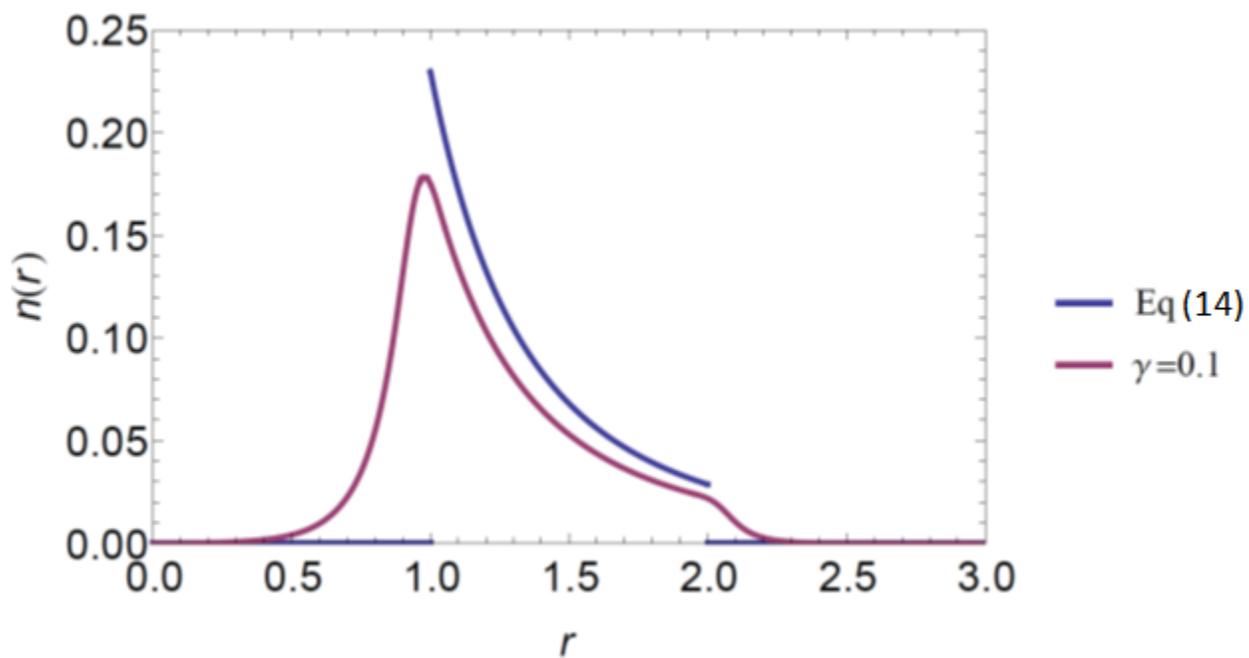

Fig. 5: Plot of the reduced density gradient $s$ for the smoothed two-electron gedanken density of Eq. (A4) vs. distance $r$ from the origin.

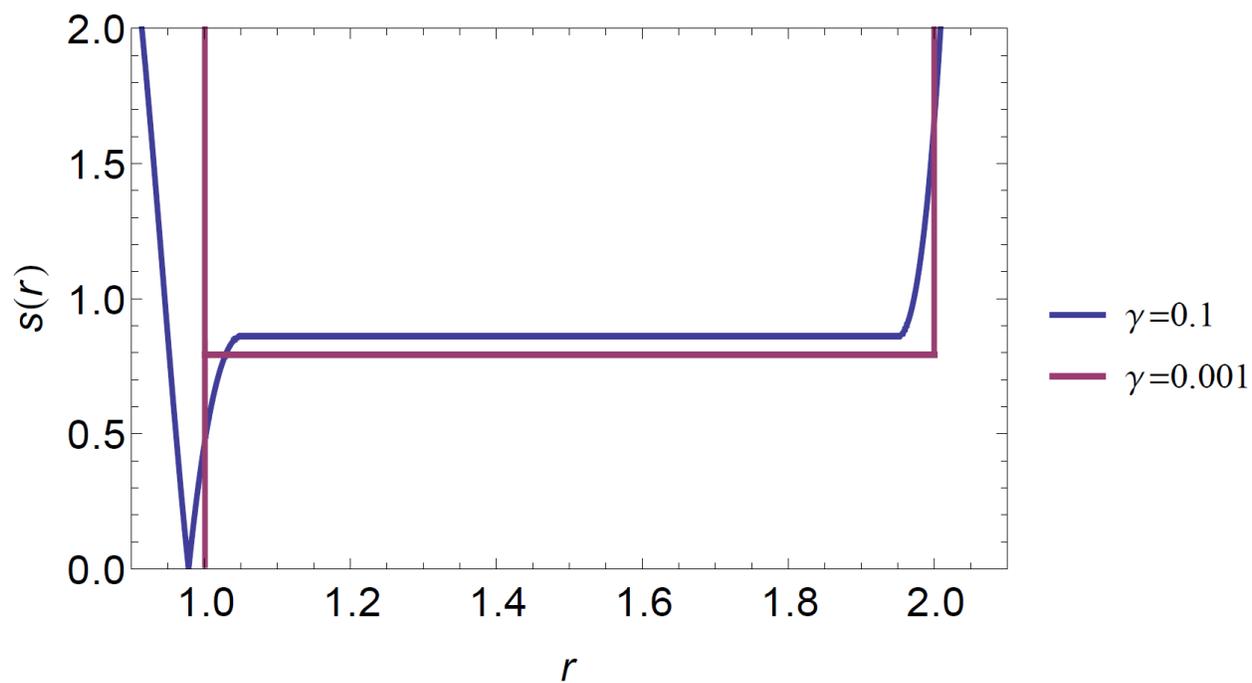

Fig. 6: Plot of the Kohn-Sham potential $v_s = \varepsilon + (1/2)\nabla^2 n^{1/2}/n^{1/2}$ for the smoothed two-electron gedanken density of Eq. (A4) vs. distance $r$ from the origin. For the gedanken density of Eq, (14), this reduces to $v_s = \varepsilon + 3/(8r^2)$ in the range $R_0 < r < R_1$. Here $\varepsilon = -1/(8\gamma^2)$ is evaluated only for $\gamma = 0.1$, to make the two curves plottable on the same scale.

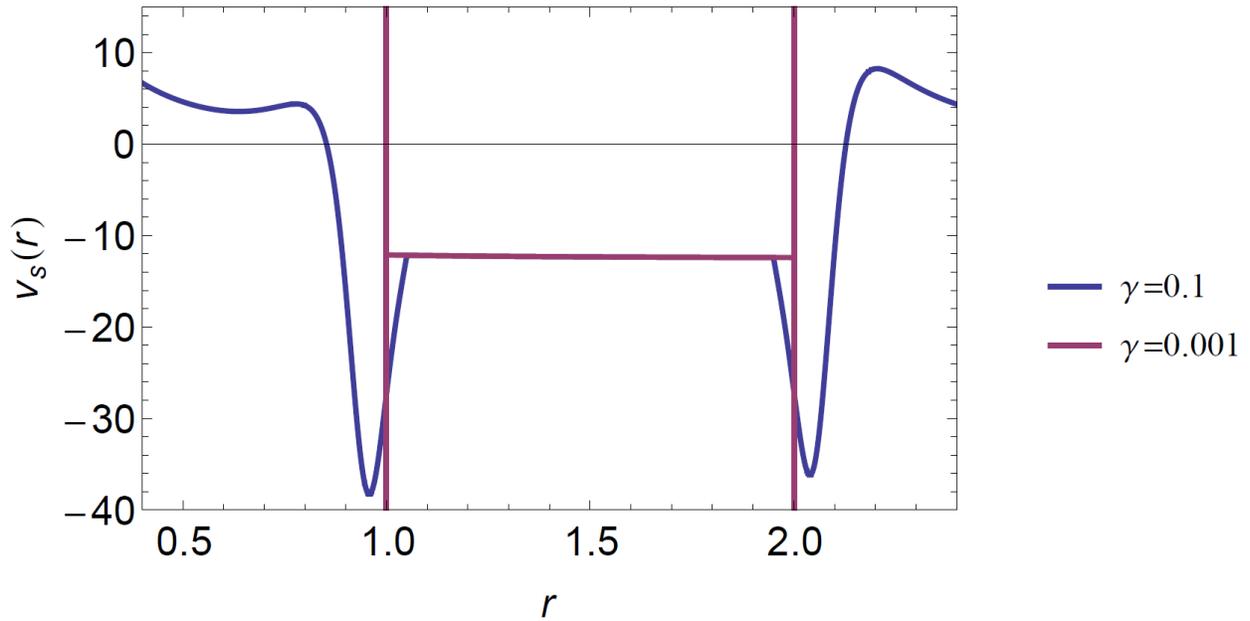